\RequirePackage{lineno}
\documentclass[twocolumn,letterpaper,aps,prc,superscriptaddress,showpacs,floatfix]{revtex4-1}

\usepackage{graphicx}
\usepackage{dcolumn}
\usepackage{bm}
\usepackage{xspace}	
\usepackage{color}

\newcommand{\pt}{\mbox{$p_T$}\xspace}

\newcommand{\sqsn}{\mbox{$\sqrt{s_{_{NN}}}$}\xspace}
\newcommand{\dau}{\mbox{$d$+Au}\xspace}
\newcommand{\dpb}{\mbox{$d$+Pb}\xspace}
\newcommand{\pau}{\mbox{$p$+Au}\xspace}

\newcommand{\hau}{\mbox{$^3\text{He}$+Au}\xspace}
\newcommand{\pp}{\mbox{$p$+$p$}\xspace}
\newcommand{\ppb}{\mbox{$p$+Pb}\xspace}
\newcommand{\pa}{\mbox{$p+A$}\xspace}
\newcommand{\da}{\mbox{$d+A$}\xspace}

\begin{document}

\title{Exploring the Beam Energy Dependence of Flow-Like Signatures in Small System $d+$Au Collisions}

\author{J.D. Orjuela Koop, R. Belmont, P. Yin, J.L. Nagle}
\affiliation{University of Colorado Boulder, Boulder, Colorado 80309, USA}

\date{\today}

\begin{abstract}
Recent analyses of small collision systems, namely \pp and \ppb at the LHC, and \pau, \dau and \hau at RHIC, have revealed azimuthal momentum anisotropies commonly associated with collective flow in larger systems. Viscous hydrodynamics and parton cascade calculations have proven successful at describing some flow-like observables in these systems.  These two classes of calculations also confirm these observables to be directly related to the initial geometry of the created system.  
Describing data at the highest RHIC and LHC energies requires a quark-gluon plasma or partonic rescattering stage, which
raises the question of how small and low in energy can one push the system before only hadronic interactions are required
for a full description.
Hence, a beam energy scan of small systems---that amounts to varying the initial temperature and the lifetime of the medium---can provide valuable information to shed light on these issues. In this paper, we present predictions from viscous hydrodynamics (\textsc{sonic} and \textsc{supersonic}), and partonic (\textsc{ampt}) and hadronic (\textsc{urmqd}) cascade calculations for elliptic ($v_2$) and triangular ($v_3$) anisotropy coefficients in \dau collisions at \sqsn = 7.7, 20, 39, 62.4 200 GeV and 5.02 TeV.
\end{abstract}

\pacs{25.75.Gz, 25.75.Gz.Ld}
\maketitle


\section{\label{sec:introduction}Introduction}

A major achievement in the field of ultrarelativistic heavy ion collisions has been the discovery that nuclear matter in its deconfined state above the transition temperature---also known as the Quark-Gluon Plasma (QGP)---behaves as a locally equilibrated, strongly coupled fluid~\cite{Snellings:2011sz}. Evidence for this claim comes from the azimuthal momentum anisotropy of emitted particles with respect to the event plane of the collision $\Psi_{n}$, as quantified by flow coefficients $v_{n}$,
\begin{equation}
\label{eq:vn}
v_{n} = \langle \cos[n(\phi-\Psi_n)] \rangle.
\end{equation}
where $n=2$ ($n=3$) corresponds to elliptic (triangular) flow coefficients.
Nearly inviscid hydrodynamics has proven to be a powerful theoretical framework to understand the physics of the medium following thermalization~\cite{Luzum:2008cw}.  It can explain the dependence of $v_n$ on particle transverse momentum \pt, particle species, collision centrality, and collision energy. However, despite great success in understanding the bulk dynamics of the QGP, the question of its formation remains open, and stronger experimental constraints on pre-equilibrium dynamics and alternative dynamical 
scenarios are needed.

To that end, the Relativistic Heavy Ion Collider (RHIC) and the Large Hadron Collider (LHC) pursued a small systems program (i.e., \dau and \ppb respectively), originally aimed at elucidating the impact of cold nuclear matter effects on QGP formation. However, it was discovered that these systems~\cite{adare_measurement_2014,alice_long_2013,atlas_observation_2012,cms_observation_2012,Adamczyk:2014fcx}, including most recently $p+p$ collisions at \sqsn = 13 TeV~\cite{Aad:2015gqa,Khachatryan:2015lva}, also exhibit azimuthal anisotropies reminiscent of those in A$+$A collisions. Namely, a near-side ($\Delta \phi \approx 0$) enhancement in two particle correlations is visible even when imposing a large pseudorapidity separation to minimize contributions to the signal from jet fragmentation and other non-flow sources.

This observation challenges the conventional understanding of how big a droplet of matter is required to form a QGP. It also raises the question of whether---and under what conditions---a hydrodynamic description is applicable in this new regime. Hence, further studies were carried out at RHIC using \pau and \hau to disentangle initial-state effects, pre-equilibrium dynamics, and final-state effects in the development of the observed flow-like signals, by leveraging the distinct intrinsic elliptic (\dau) and triangular (\hau) geometry of these systems~\cite{nagle_exploiting_2013,Adare:2015ita}.

Studies showed that nearly inviscid hydrodynamics, as implemented in the \textsc{sonic} model~\cite{Habich:2014jna}, provides a good description of the measured $v_2$ in \dau, \hau, and most recently in \pau~\cite{itaruqm2015}, as well as $v_3$ in \hau~\cite{Adare:2015ita}. Furthermore, including a pre-equilibrium flow stage, as present in the \textsc{supersonic} model~\cite{Romatschke:2015gxa}, results in a closer agreement with experimental data, with the largest influence in triangular flow.

However, it was also shown that completely different physics, as encoded in A-Multi-Phase-Transport (\textsc{ampt}) model with a modest parton-parton cross section, can reproduce the observed two-particle correlations~\cite{ma_long-range_2014} and anisotropy coefficients~\cite{bzdak_elliptic_2014} in \ppb and \pp at the LHC, as well as the measured $v_2$ and $v_3$ for $p_T < 1$ GeV/c in \pau, \dau, and \hau at RHIC~\cite{Koop:2015wea}. In this context, the physics responsible for the development of azimuthal momentum anisotropy---although not yet fully understood---has been identified to be a combination of rather limited partonic scattering and freeze out~\cite{He:2015hfa}, and a simple coordinate space coalescence followed by hadronic scattering~\cite{Koop:2015wea}.

The Ultra-Relativistic Quantum Molecular Dynamics (\textsc{urqmd}) model is another theoretical framework that has been utilized to understand flow in small systems~\cite{Zhou:2015iba}, although it typically does not describe $v_n$ values for large collision systems at \sqsn = 200 GeV~\cite{PhysRevC.72.064911}.   Since \textsc{urqmd} does not include any partonic or QGP stage, it is an excellent comparison model to the above calculations.

Although the above studies have provided substantial insight into the role of initial geometry in small system anisotropies, further study is required to fully understand how the azimuthal anisotropy accrues in these short-lived systems, since they do not live long enough to fully translate initial-state geometry to final-state momentum anisotropy. Pre-equilibrium dynamics as well as later stage hadronic interactions are expected to play a larger role than in A$+$A collisions~\cite{Romatschke:2015gxa,Zhou:2015iba}. To that end, the RHIC program is planning to include a small system beam energy scan in its 2016 run period, colliding \dau at a variety of center-of-mass energies from \sqsn = 20 to 200 GeV.

In this paper, we utilize the theoretical frameworks---\textsc{(super)sonic} and \textsc{ampt}---that provide a reasonable agreement with currently available datasets, to make predictions for elliptic and triangular Fourier coefficients, $v_2$ and $v_3$, for \dau at \sqsn = 7.7, 20, 39, 62.4, and 200 GeV, as well as \dpb at \sqsn = 5.02 TeV. We also present $v_2$ results from \textsc{urqmd}  at these energies. 
We begin by describing the \textsc{ampt}, \textsc{urqmd}, and \textsc{(super)sonic} models, as well as the methodology used to compute $v_n$. We then present and compare the results for all energies obtained with the three models. Finally, we discuss the results and draw conclusions from them. 

\section{\label{sec:methods}Model Calculations}
\subsection{\textsc{AMPT}}
The \textsc{ampt} generator~\cite{lin_multiphase_2005} has been established as a useful tool in the study of heavy ion collisions. In \textsc{ampt}, initial conditions are generated using a Monte Carlo Glauber model as implemented in the \textsc{hijing} generator~\cite{wang_hijing_1994}. Zhang's Parton Cascade (\textsc{zpc}) is then used to model partonic scattering, with a parton-parton interaction cross section inversely proportional to the parton screening mass. Hadronization is implemented as a simple coordinate space coalescence, where the closest particle pairs and triplets are bound into mesons and baryons, respectively. Lastly, the model incorporates a hadronic scattering phase using A-Relativistic-Transport (\textsc{art}).   

In this paper, we follow the methodology of Ref.~\cite{Koop:2015wea} and use \textsc{ampt} Version 2.26 with string melting turned on.  
It is notable that in the string melting case, all partons are quarks and anti-quarks in the exact number to provide the constituent quarks
of the final state hadrons via coalescence.   A momentum dependent formation time for these (anti)quarks is implemented, effectively
giving them a brief free streaming time that generates an initial space-momentum correlation.   This mechanism plays an important role within \textsc{ampt}
on the final momentum anisotropies.

We have modified the default \textsc{ampt} Monte Carlo Glauber to sample the positions of the deuteron nucleons from the Hulth\'en wavefunction description of the nucleus. Additonally, we modified the initial nucleon interactions from the default
\textsc{hijing} implementation to one where the nucleon-nucleon cross section depends exclusively on the energy of the collision, following Ref.~\cite{Miller:2007ri,Loizides:2014vua}. The values of the cross section corresponding to the center-of-mass energies comprising this study are presented in Table~\ref{tab_crosssection}.   This modification was implemented so that the \textsc{ampt} and hydrodynamic calculations would start with identical initial geometries.
\begin{table}[tbh]
\caption{\label{tab_crosssection} Nucleon-nucleon cross section as a function of collision energy in the Monte Carlo Glauber model.}
\begin{tabular}{c|c}
\hline
\sqsn [GeV] & $\sigma_{NN}$ [mb] \\
\hline 
7.7 & 31.2 \\ 
20.0 & 32.5 \\
39.0 & 34.3 \\ 
62.4 & 36.0 \\ 
200.0 & 42.3 \\
5020.0 & 60.3 \\
\hline
\end{tabular}
\end{table}
We ran approximately 4 million central \dau \textsc{ampt} events at the highest energy and 8 million at the lowest energy, with impact parameter $b<2$ fm.  We have used the same
geometry selection for all model calculations, so they are directly comparable, even though later
comparison with experimental \dau energy scan data may require further matching to experiment-specific centrality selections.  
In the case of \textsc{ampt}, the matching is straightforward in that one can select on multiplicity or energy
in the same pseudorapidity as the experiments and determine comparable centrality quantiles---as done for example in Ref.~\cite{Adare:2015lcd}.

It has been shown that a partonic cross section of $\sigma_{part} = 1.5$ mb can reasonably reproduce the measured $v_2$ up to \pt $\approx 1$ GeV/c in \dau and \hau at RHIC energies~\cite{Koop:2015wea}. Hence, in keeping with these studies, we use the identical partonic cross section for all collision energies. It is important to note that our assumption of a constant cross section, independent of the collision energy, has been warranted by studies demonstrating how the same value of $\sigma_{part} = 1.5$ mb can reproduce flow observables in \pp and \ppb at LHC energies~\cite{ma_long-range_2014}.

For every \textsc{ampt} event, we access the initial coordinates $(r_i,\phi_i)$ of the participant nucleons and smear them by a Gaussian of $\sigma=0.4$ fm. We then compute the second and third order participant plane as follows
\begin{equation}
\Psi_n = \frac{\text{arctan}[\langle r^2\sin(n\phi) \rangle/ \langle r^2\cos(n\phi) \rangle]}{n} + \frac{\pi}{n}.
\end{equation}
We calculate $v_2$ and $v_3$ for unidentified charged hadrons at midrapidity $|\eta|<2$ according to Eq. (\ref{eq:vn}). Calculating anisotropy moments in this manner has the advantage of excluding non-flow effects including resonance decays and jet fragmentation.

\begin{figure*}[hbtp]
\centering
\includegraphics[scale = 0.85]{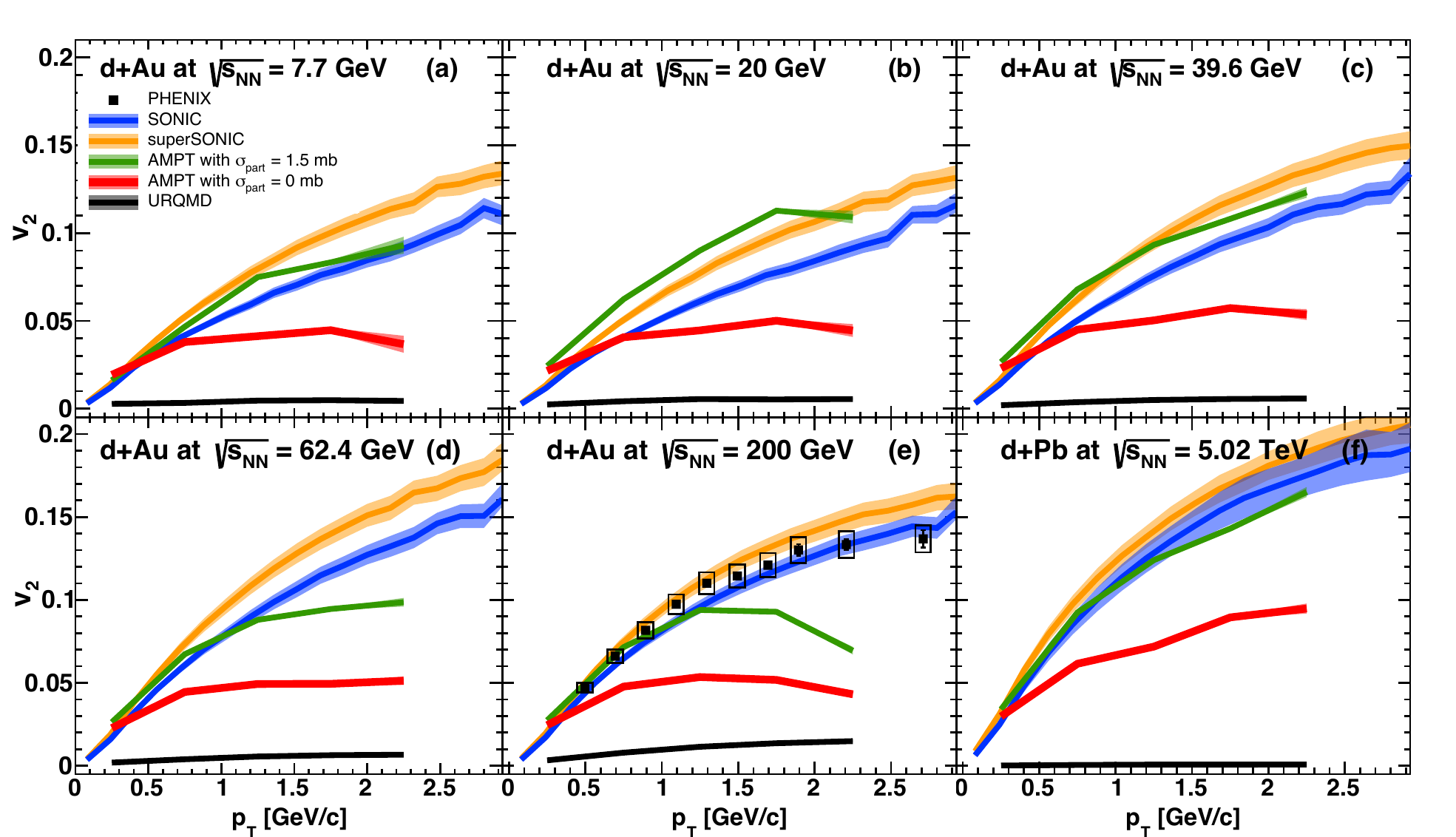}
\caption{Transverse momentum dependence of $v_2$ in \dau collisions at $\sqsn = 7.7,~20.0,~39.6,~62.4,~200.0$ GeV and in \dpb collisions at \sqsn = 5.02 TeV. Results are shown from viscous
hydrodynamic calculations (\textsc{sonic} and with pre-equilibrium dynamics \textsc{supersonic}) as well as the parton cascade model \textsc{ampt} (with and without
a partonic scattering stage), and the purely hadronic cascade model (\textsc{urqmd}).  Published experimental data for $v_2$ in \dau at $\sqsn = 200$ GeV is also shown.}
\label{fig_v2_all}
\end{figure*}
\begin{figure*}[hbtp]
\centering
\includegraphics[scale = 0.85]{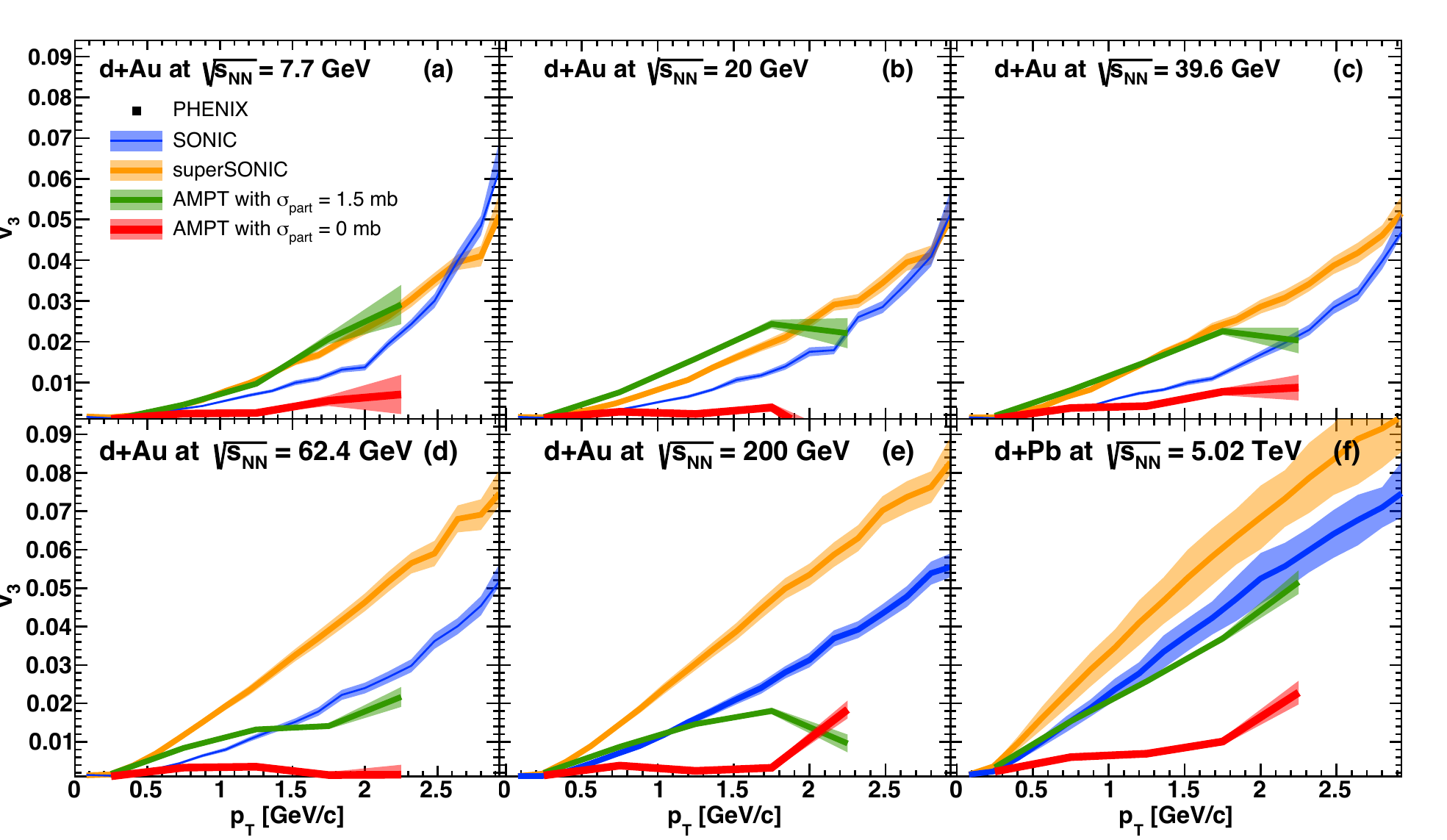}
\caption{Transverse momentum dependence of $v_3$ in \dau collisions at $\sqsn = 7.7,~20.0,~39.6,~62.4,~200.0$ GeV and in \dpb collisions at \sqsn = 5.02 TeV. Results are shown from viscous
hydrodynamic calculations (\textsc{sonic} and with pre-equilibrium dynamics \textsc{supersonic}) as well as the parton cascade model \textsc{ampt} (with and without
a partonic scattering stage).}
\label{fig_v3_all}
\end{figure*}
\clearpage

We have also run \textsc{ampt} with the partonic cross section set to zero ($\sigma_{part} = 0.0$ mb).    In this case, one can 
isolate the effects of hadronization and the subsequent hadron casade on the development of $v_{n}$.   One can hypothesize that  at some lower collision energy for small systems there is no partonic stage.  In that case, experimental results might move from
agreement with \textsc{ampt} with $\sigma_{part} = 1.5$ mb at higher energy, towards the \textsc{ampt} with $\sigma_{part} = 0.0$ mb (i.e. with no partonic stage) at lower energy.  

\subsection{\textsc{URQMD}}

In order to further check expectations that for small systems at lower energies the medium is purely hadronic, we have also employed the \textsc{urqmd} model.
The \textsc{urqmd} model Version 3.4~\cite{Bass:1998ca} is a hadronic transport model where initial nucleon positions are determined from a Woods-Saxon distribution. Hadrons, predominantly arising from strings are propagated via relativistic Boltzmann transport. As with \textsc{ampt}, we modified the \textsc{urqmd} code to sample deuteron nucleon coordinates from the appropriate Hulth\'en wave function of the nucleus. 
The \textsc{urqmd} cutoff time for interactions was set to $t_{cutoff} = 20$ fm/c.
Since the \textsc{urqmd} model was originally designed for the study of heavy ion collisions at energies between those of SIS and RHIC, the default model does not apply to LHC energies. In that case, care must be taken to compile the source code in LHC mode prior to running any calculation.

It is interesting to compare the results from \textsc{ampt} with no partonic scattering to those from \textsc{urqmd}. In \textsc{ampt}, as detailed above,
the use of a standard Monte Carlo Glauber allows for a straightforward determination of the participant nucleons and the subsequent calculation of $v_n$ relative to their symmetry plane, thus excluding any possible non-flow effects from the measured signal. On the other hand in \textsc{urqmd}, although initial nucleon coordinates are sampled from an appropriate description of the colliding nuclei, participants are not determined from the initial impact, but rather from the evolution of the system, as nucleons are struck over time by other nucleons or---more commonly---by the products of previous inelastic collisions. This results in an ambiguity in the definition of the \textsc{urqmd} participant plane.   Through careful study, we find that a reliable participant plane $\Psi_{n}$ 
can be extracted by including only nucleons struck by other nucleons or by restricting the interactions to occur at early times.   

The \textsc{urqmd} calculations in Ref.~\cite{Zhou:2015iba} use the two-particle cumulant method~\cite{Bilandzic:2010jr} with an
imposed pseudorapidity gap of $|\Delta \eta| > 0.2$ and $> 0.8$.   In order to cross check their results, we have implemented the two-particle cumulant method with varying pseudorapidity gaps.    In addition, in order to extract $v_n$ from \textsc{urqmd} minimizing non-flow contributions, we implemented four-particle cumulants following Ref.~\cite{Bilandzic:2010jr}. It was found that $c_2\{4\} > 0$, resulting in a complex normalization for $v_2$. It has been reported that this is a limitation of the method when dealing with large fluctuations in low multiplicity events~\cite{PhysRevC.72.064911}. 


\subsection{Hydrodynamics}

In the viscous hydrodynamic \textsc{sonic} model, initial conditions are generated for each event using the identical Monte Carlo Glauber procedure described above for \textsc{ampt}. These conditions are then translated into fluid cells propagated with 2+1 dimensional conformal viscous hydrodynamics as described in Ref.~\cite{Habich:2014jna}. Lastly, final-state interactions are modeled with the hadronic cascade code \textsc{B3D}~\cite{Novak:2013bqa}. 

The \textsc{sonic} implementation of viscous hydrodynamics has been successfully applied to A$+$A collisions at RHIC. However, it incorporates no pre-equilibrium dynamics. These are implemented in an extension of the model, known as \textsc{supersonic}~\cite{Romatschke:2015gxa}. In the latter, the AdS/CFT correspondence is used to derive a relation between the gradient of the initial energy density distribution and the radial dependence of fluid cell velocities, which accounts for the presence of pre-equilibrium dynamics. The final $v_n(\pt)$  are obtained by summing over final state particles, using the formalism described in Ref.~\cite{Romatschke:2015gxa}.

\section{\label{sec:results}Results and Discussion}

Figure~\ref{fig_v2_all} presents a full suite of theoretical predictions for elliptic anisotropy coefficients $v_2$ as a function of \pt for \dau collisions at \sqsn = 7.7, 20.0, 39.0, 62.4, 200.0 GeV and 5.02 TeV.    Experimental data in \dau at \sqsn = 200 GeV for $v_2$ measured by the PHENIX collaboration is shown in panel (e).   Similar predictions for $v_{3}$ are shown in Figure~\ref{fig_v3_all}.   We now discuss each category of predictions and possible implications
for future measurements.

\subsection{Hydrodynamics}

In the context of hydrodynamics, there is a well-understood physical picture of how initial inhomogeneities in the energy density deposition due to intrinsic geometry or fluctuations are translated into final-state momentum anisotropies. This model has been successfully applied to large collision systems. In such cases, the duration of the QGP phase is long enough to bring the $v_n$ signal from the hydrodynamic phase to saturation and dominate over other sources of flow prior to thermalization or following hadronization. However, it is precisely the dominance of the hydrodynamic phase that needs to be revisited for small collision systems.

Results from hydrodynamics are shown as the blue (\textsc{sonic}) and yellow (\textsc{supersonic}) curves.   The systematic uncertainty
bands are calculated from the sensitivity to the viscous second order corrections as detailed in Ref.~\cite{Romatschke:2015gxa}.   Both calculation results for \dau at \sqsn = 200 GeV are in reasonable agreeement with the experimental data.   
The additional effect on $v_2$ of pre-equilibrium flow in \textsc{supersonic} of +10\% relative to \textsc{sonic} cannot be discriminated
with the current experimental and theoretical uncertainties.  

\begin{figure}
\centering
\includegraphics[scale = 0.4]{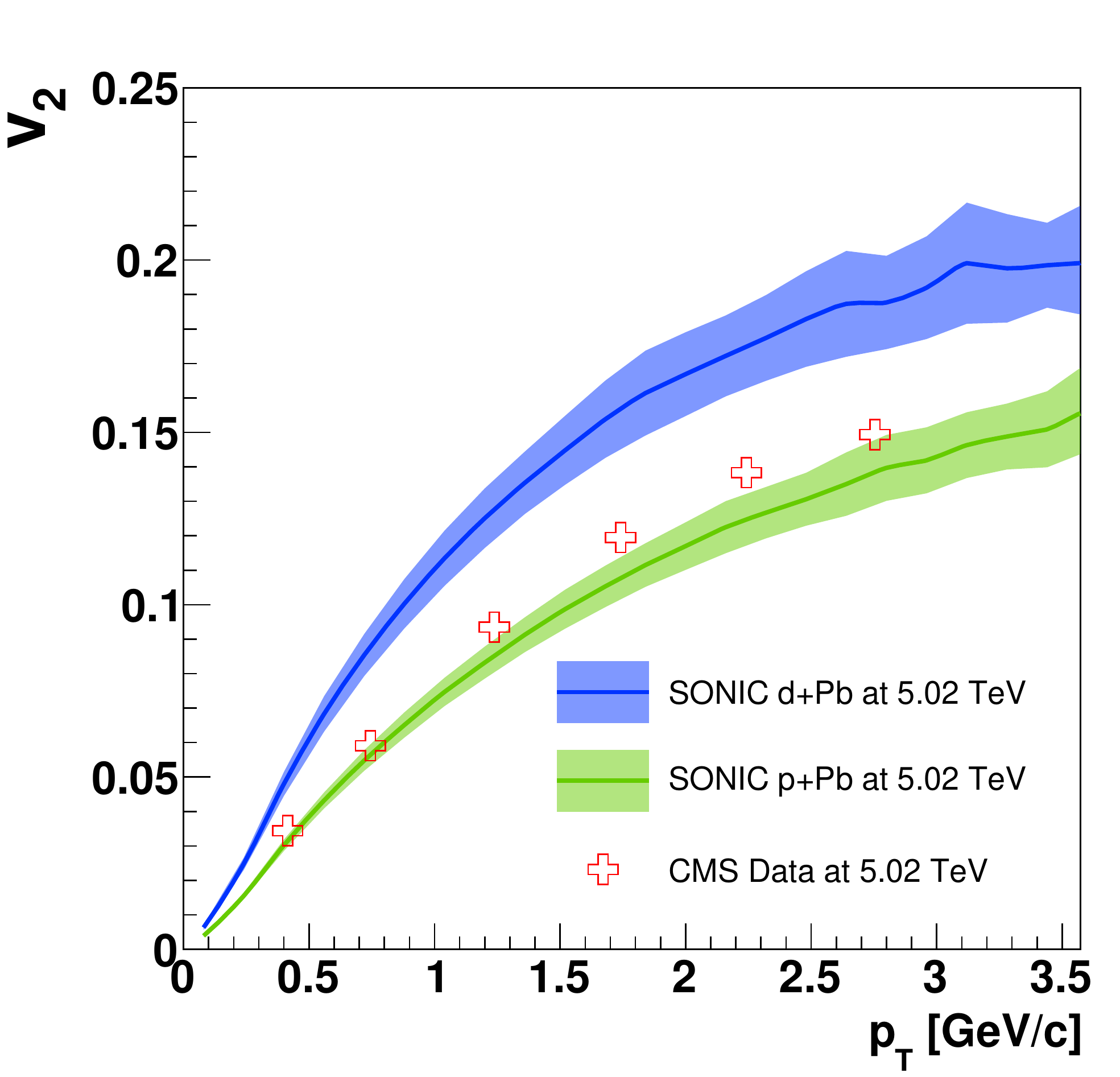}
\caption{Transverse momentum dependence of $v_2$ in \ppb and \dpb at \sqsn = 5.02 TeV from the \textsc{sonic} hydrodynamic model. CMS measurements in central \ppb events are shown for comparison.}
\label{fig_ppb_comp}
\end{figure}

At the LHC, there is experimental data in \ppb at \sqsn = 5.02 TeV~\cite{Chatrchyan:2013nka} 
and a comparison with the \textsc{sonic} calculation is shown in Figure~\ref{fig_ppb_comp}.    We observe a reasonable agreement between the hydrodynamic
calculation and data, validating the model at LHC energies. Furthermore, we see that \textsc{sonic} predicts a substantially larger $v_2$ in \dpb compared
with \ppb collisions.   A major ambiguity in the interpretation
of LHC \pp and \ppb results is the widely differing models of the initial geometry and their event-wise fluctuations.   In contrast, in \dpb collisions the initial geometry is mostly determined by the positions of the two nucleons in the deuteron rather than in modeling the proton substructure and shape fluctuations.   
Specifically, the Monte Carlo Glauber calculated eccentricity and the IP-Glasma calculated eccentricity~\cite{Schenke:2014gaa} differ by less than 10\% in
\da collisions and differ by more than 125\% in \pa collisions.   Recently it has been suggested that the IP-Glasma calculation may be augmented by
including shape fluctuations of the proton (not just size fluctuations)~\cite{Schlichting:2014ipa}, adding another degree of freedom to the calculation.    In contrast, with \dpb data one can pin down the medium properties since the initial conditions are much better constrained, and then use the fixed medium properties to better understand the initial conditions in \ppb. Thus, despite technical obstacles, future \dpb running at the LHC should be seriously considered.

\begin{figure*}[hbtp]
\centering
\includegraphics[scale = 0.85]{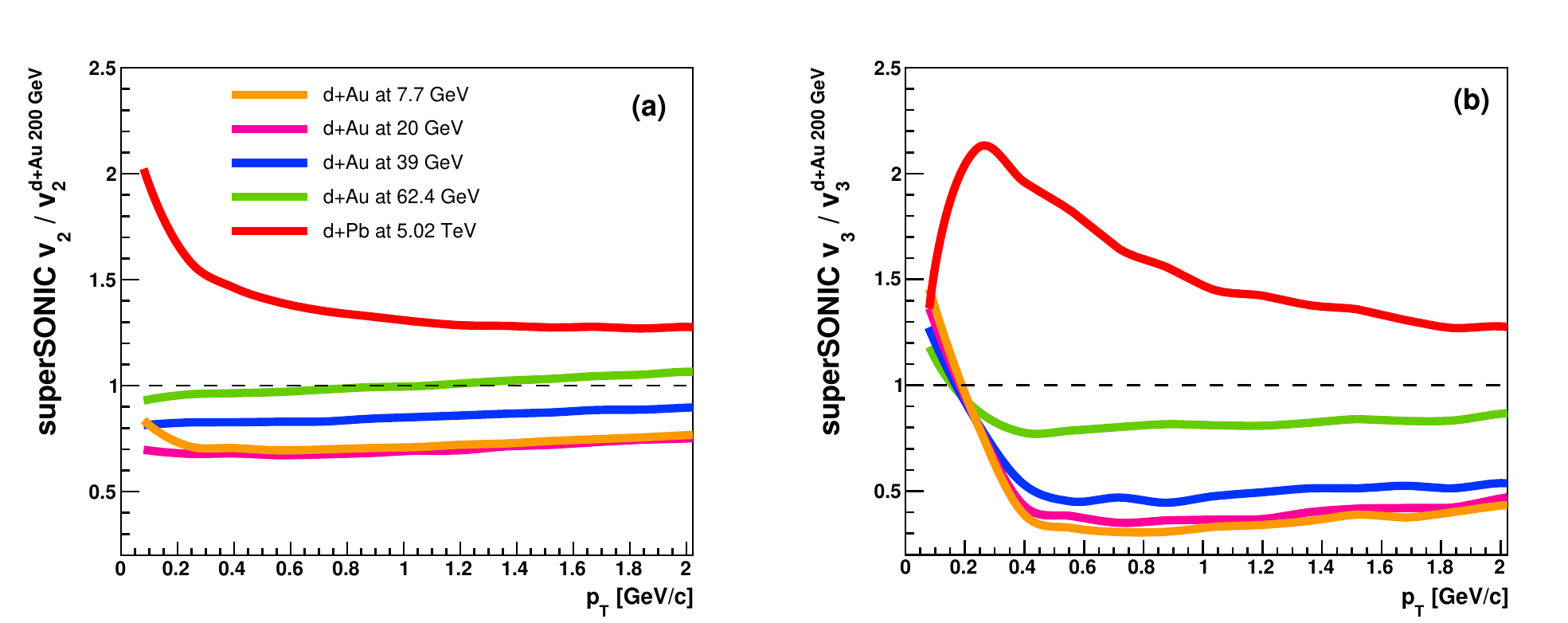}
\caption{
Panels (a) and (b) show the ratio of $v_2$ ($v_3$) as a function of \pt at different energies to \dau $v_2$ at \sqsn = 200 GeV from the \textsc{supersonic} hydrodynamic calculation.}
\label{fig_v2_v3_energyratio_supersonic}
\end{figure*}

The calculation results for \dau collisions at lower energies 7.7 - 62.4 GeV show a similar rise in the $v_{2}$ with \pt, though with modestly reduced magnitude.  Shown in Figure~\ref{fig_v2_v3_energyratio_supersonic}, panel (a) are the ratios of $v_{2}$ as a function of \pt for each energy to the value in \dau at \sqsn = 200 GeV.   The $v_{2}$ 
magnitude decreases monotonically from the highest energy to the lowest \sqsn with only modest \pt dependence.   It is striking that in going from the LHC
energy of 5.02 TeV to RHIC top energy of 200 GeV, there is an approximate 25-30\% decrease in the $v_{2}$ and then between 200 GeV and the lowest
energy of 7.7 GeV the decrease is only another 25-30\%.     Given that large decrease in the energy available to be deposited to create the hot fireball, 
why is there no strong collapse of the $v_2$ at the lower energies?

\begin{figure}[hbtp]
\centering
\includegraphics[width=0.5\textwidth]{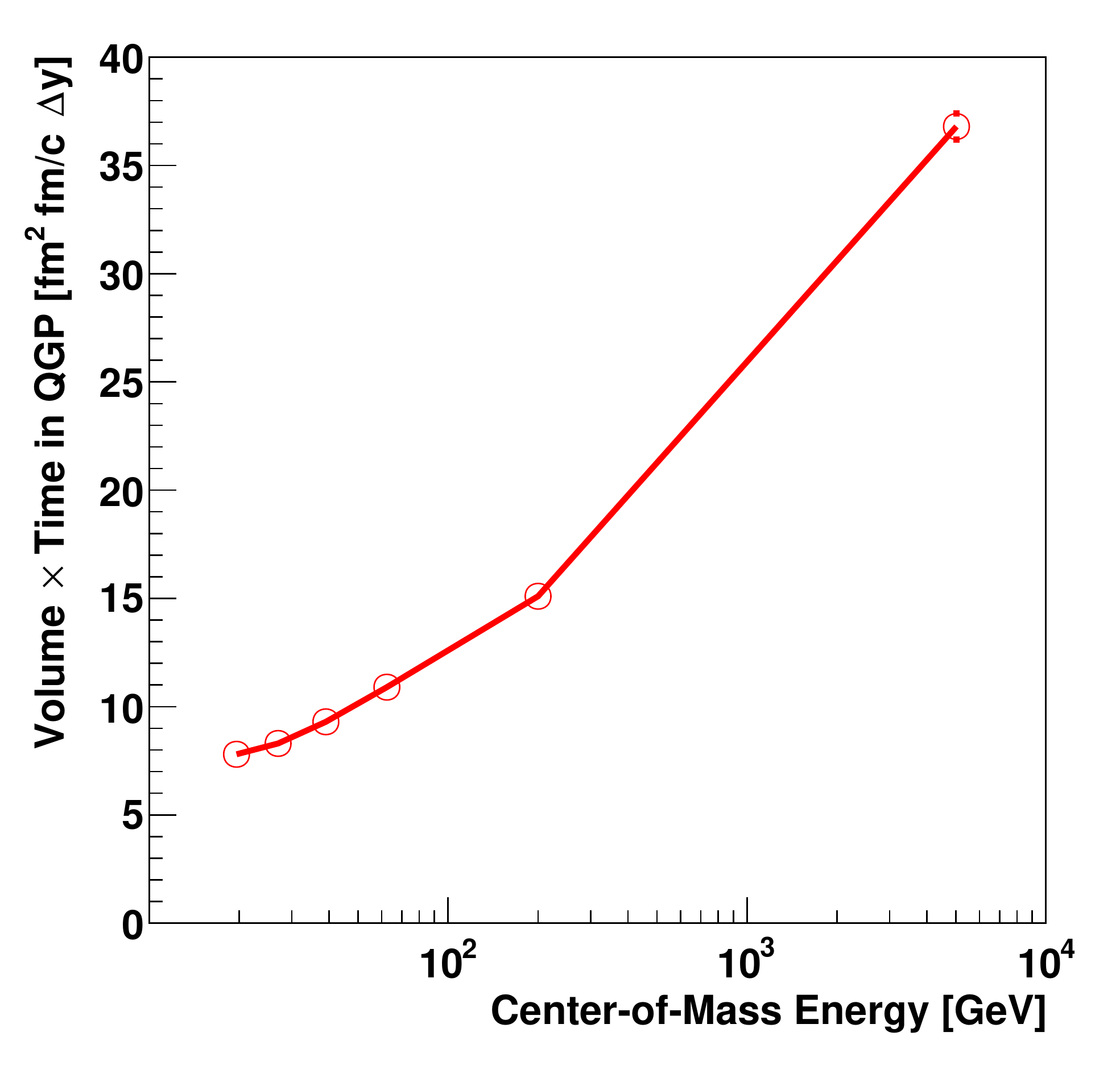}
\caption{Collision energy dependence of the summed space-time volume element in the QGP in viscous hydrodynamics.}
\label{fig_spacetime}
\end{figure}

To shed further light on this question, we sum the space-time volume of all fluid elements hotter than the transition temperature for every time step in our hydrodynamic simulations. The result, plotted as a function of collision energy, is shown in Figure~\ref{fig_spacetime}. We observe a five-fold increase in the summed volume between the lowest and highest energies. Even at \sqsn = 7.7 GeV there is a space-time volume of QGP of order 8 fm$^2$~$\Delta y$ ~fm/c.   As a very rough picture, one can think of this as a 2 fm $\times$ 2 fm transverse area of QGP that lasts for a time 2 fm$/c$ with rapidity interval $\Delta y$.
Therefore, within the viscous hydrodynamics framework, the answer is that there is no minimum size to the QGP droplet. This makes sense since there is no expected first order phase transition with bubbles, but rather a continuous change where the space-time volume of the QGP simply decreases, playing a smaller and smaller role.  Thus, even a rather small and short lived QGP phase is enough to quickly generate significant elliptic flow.

Figure~\ref{fig_v2_v3_energyratio_supersonic}, panel (b) shows the ratios of $v_{3}$ as a function of \pt for each energy to the value in \dau at \sqsn = 200 GeV.
The $v_3$ calculation results show a much more dramatic drop with collision energy.   The decrease in $v_{2}$ between 200 GeV and 7.7 GeV was 25-30\%, but for
$v_{3}$ the change is a reduction of roughly 300\%.
As noted in Ref.~\cite{nagle_exploiting_2013}, the triangular flow simply takes a longer time to develop (i.e.
to translate the spatial triangularity into momentum anisotropy) and thus it is natural that this signal is much more sensitive to the shorter and shorter
QGP lifetime at the lower energies.   

\begin{figure*}[hbtp]
\centering
\includegraphics[scale = 0.8]{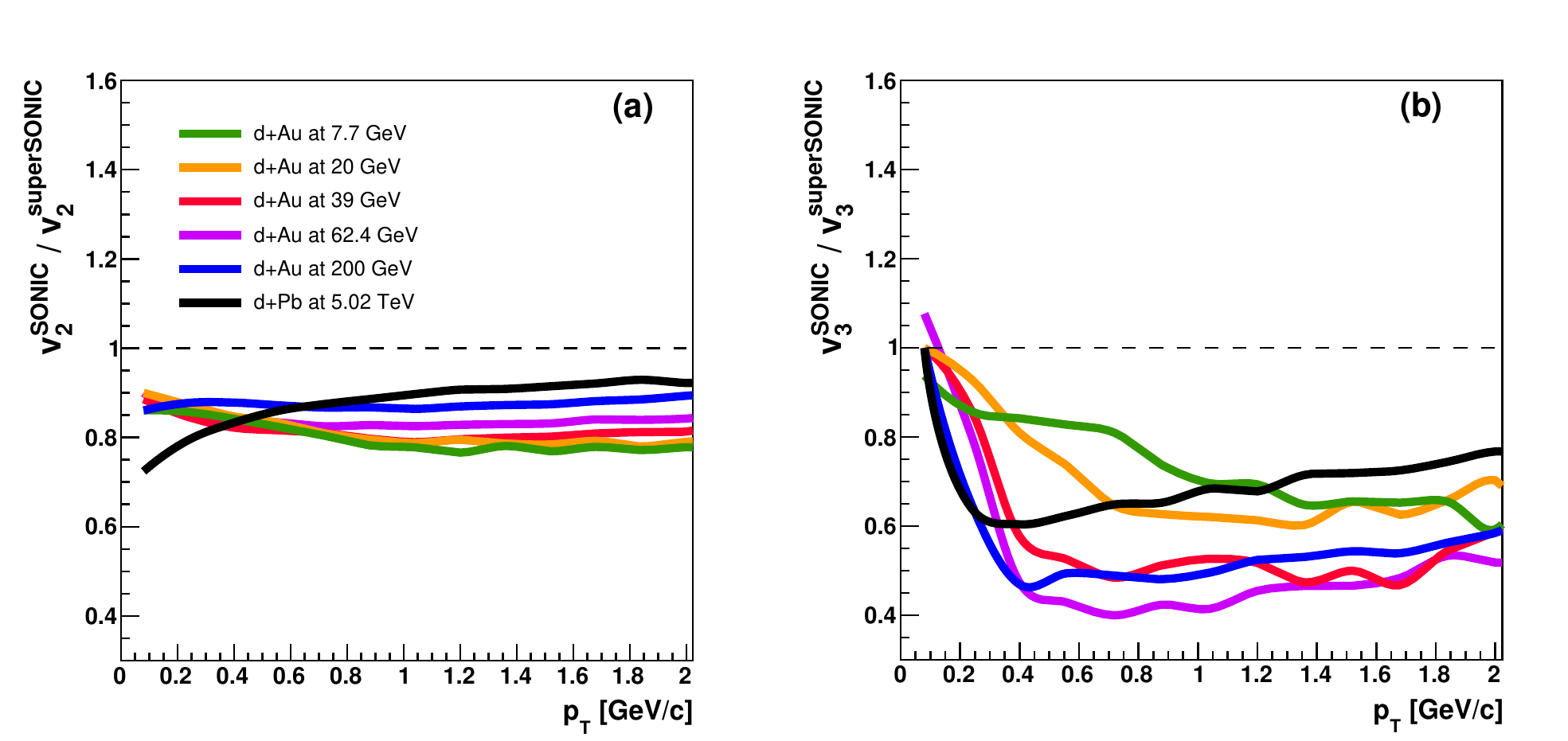}
\caption{Ratio of $v_2(p_T)$ in panel (a) and $v_3(p_T)$ in panel (b) from \textsc{sonic} to \textsc{supersonic} for \dau at all collision energies.}
\label{fig_hydro_ratio}
\end{figure*}

This effect is also quite striking when evaluating the influence of pre-equilibrium flow via the 
comparison of \textsc{sonic} and \textsc{supersonic}.   Shown in panel (a) of Figure~\ref{fig_hydro_ratio} is the ratio of $v_2$ 
computed with \textsc{sonic} to \textsc{supersonic} for all collision energies.
For $v_2$ we see that the two calculations differ the most at the lowest energy, and are closest at the highest energy, with the ratio increasing monotonically with energy. This demonstrates that as the contribution of the hydrodynamic phase to total flow becomes weaker, pre-equilibrium dynamics play an increasingly significant role.  The same ratio for $v_3$ is shown in panel (b).  Pre-equilibrium also appears to play a stronger role in going from 5.02 TeV to 62.4 GeV. However, below that energy the $v_3$ values are quite small, less than 1\% for $p_{T} < 1$ GeV/c, and the ratios no longer follow a monotonic trend.   The
best test of the influence of pre-equilibrium dynamics is with signals that take the longest time to develop, such as $v_3$.    It will also be very interesting
to compare such experimental data in these small systems where the geometry is well quantified with recent results from the beam energy dependence of
$v_3$ in peripheral Au+Au collisions reported by the STAR collaboration~\cite{Adamczyk:2016exq}.

\begin{figure*}[hbtp]
\centering
\includegraphics[scale = 0.8]{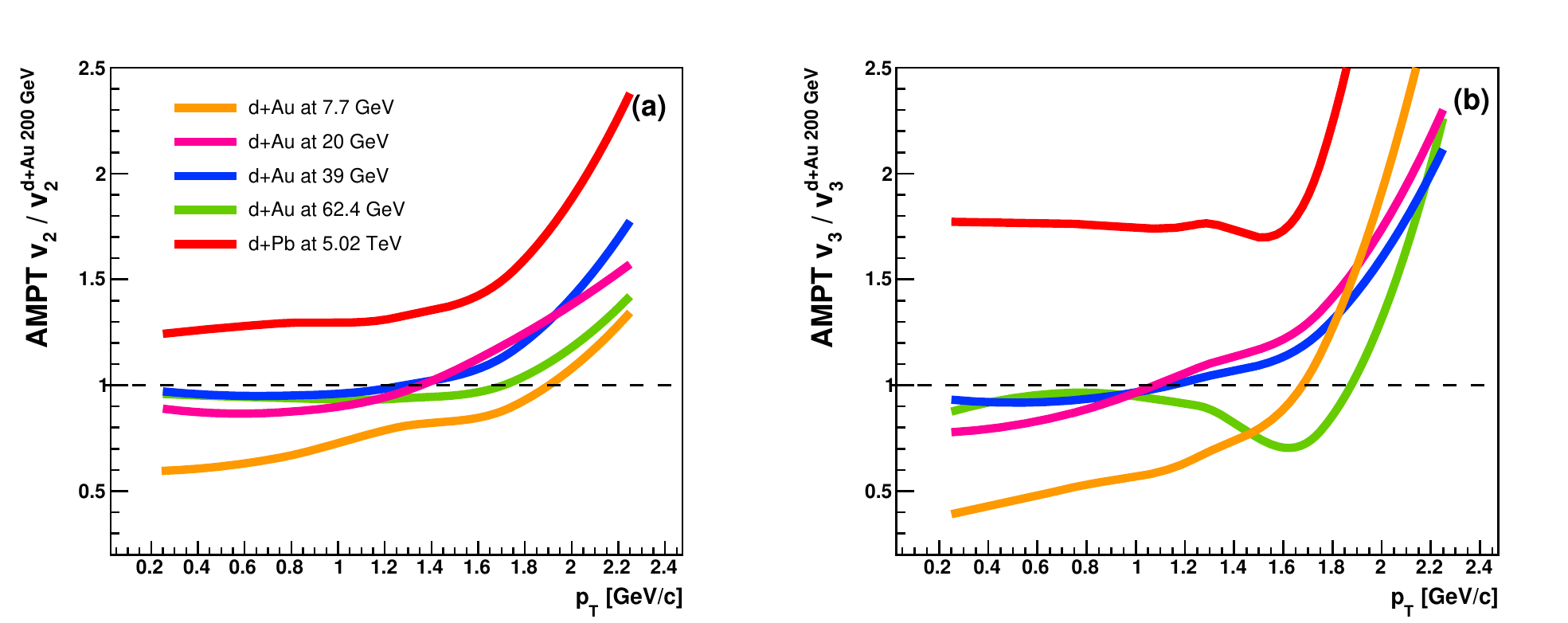}
\caption{
Panels (a) and (b) show the ratio of $v_2$ and $v_3$, respectively, as a function of \pt at different energies to \dau $v_2$ and $v_3$ at \sqsn = 200 GeV from AMPT.}
\label{fig_v2_v3_energyratio_AMPT}
\end{figure*}


\subsection{Partonic versus Hadronic Scattering}

We now discuss the results of \textsc{ampt} with and without partonic scattering, and the purely hadronic model \textsc{urqmd}.
\textsc{ampt} results with $\sigma_{part} = 1.5$ mb are shown in green, and the results with no partonic scattering $\sigma_{part} = 0$ mb are
shown in red for $v_2$ in Figure~\ref{fig_v2_all} and for $v_3$ in Figure~\ref{fig_v3_all}.
Across collision energies, calculations using \textsc{ampt} with $\sigma_{part} = 1.5$ mb yield a sizable $v_2$ that grows as a function of \pt.   
The magnitude of $v_2$ decreases modestly with decreasing collision energy.    
The \textsc{ampt} calculations with partonic scattering agree with the experimental data for \dau at 200 GeV
up to \pt $\approx 1$ GeV/c and then decrease at higher \pt.  

\begin{figure}[hbtp]
\centering
\includegraphics[scale = 0.4]{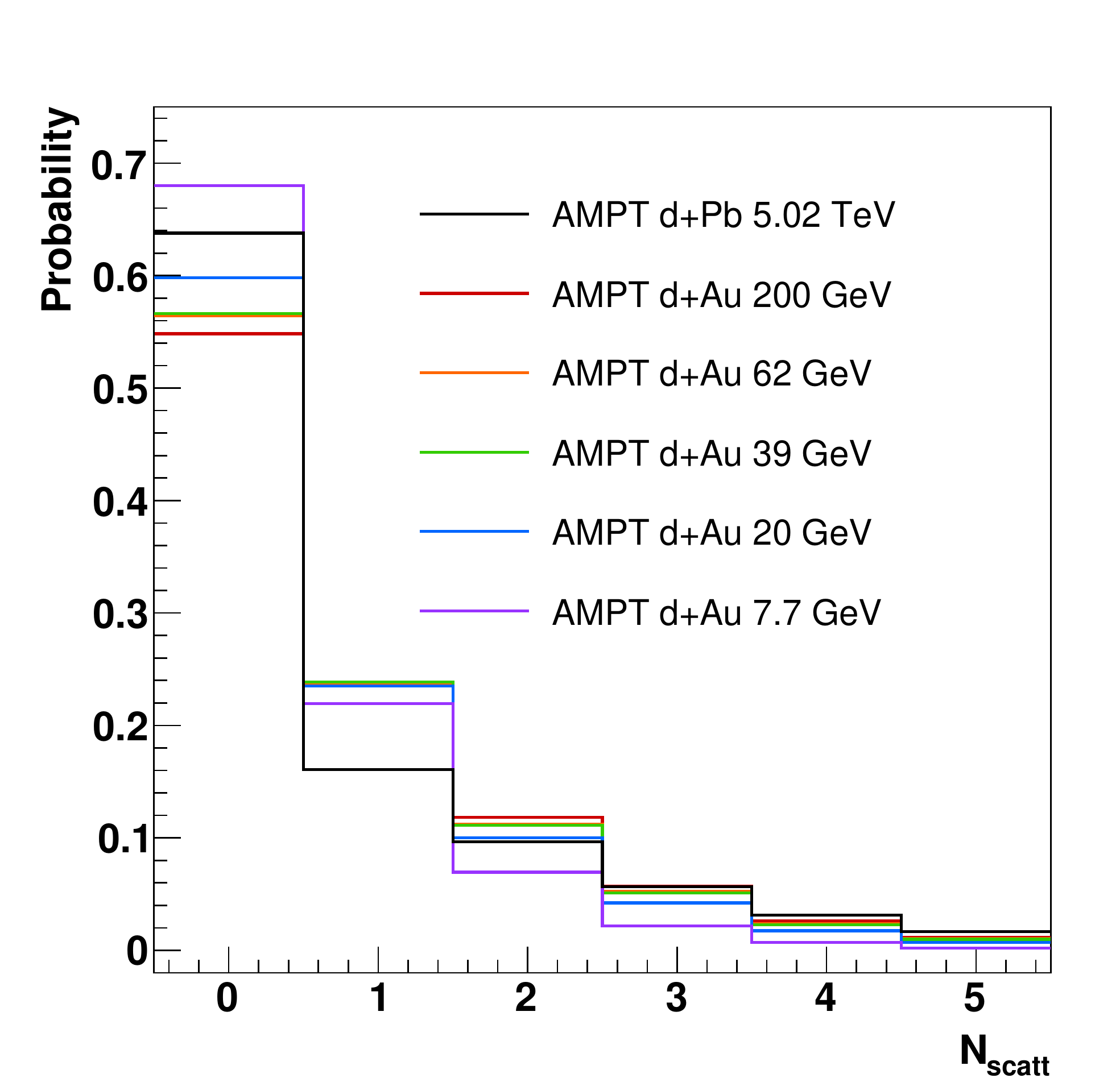}
\caption{Probability for a parton in \textsc{ampt} to undergo $N$ scattering events prior to hadronization, for \dau and \dpb collisions at a variety of collision energies.}
\label{fig_nscatt}
\end{figure}

It is striking that \textsc{ampt} can yield $v_2$ values comparable to hydrodynamics---see for example Ref.~\cite{Adare:2015ema}. Previous studies have shown that, in small systems at \sqsn = 200 GeV, incoherent partonic scattering only accounts for approximately half of the total developed $v_2$, with the remainder arising from coalescence and hadronic interactions~\cite{Koop:2015wea}.  As shown in Figure~\ref{fig_v2_v3_energyratio_AMPT}, the $v_2$ results decrease
by approximately 25-30\% from 5.02 TeV to 200 GeV, and then another 25-40\% from 200 to 20 GeV - quite similar in magnitude to the effect
in the hydrodynamic calculations.
In order to study how partonic scattering in \textsc{ampt} changes as a function of collision energy, Figure~\ref{fig_nscatt} shows the probability of a (anti)quark to undergo $N$ scattering events prior to freeze out. It is noteworthy that over 50\% of the quarks do not scatter at all, and the remainder scatter only a few times. It is seen that for \dau, the probability of not scattering increases monotonically with collision energy. However, the dependence of this effect on \sqsn is found to be small, of order 10\%. 

In order to isolate the effects of hadronization and the subsequent hadron cascade on the development of $v_n$, \textsc{ampt} was run 
with $\sigma_{part} = 0$ mb (i.e. with no partonic scattering) . 
By setting the partonic cross section to $\sigma_{part} = 0$ mb, it is possible to turn off partonic scattering, such that azimuthal anisotropies develop from string melting, coalescence, and hadronic scattering.   From Figure~\ref{fig_v2_all}, we see that this hadronic phase accounts for approximately half of the $v_2$ signal above $\pt \approx 1$ GeV/c, regardless of collision energy. However, Figure~\ref{fig_v3_all} shows that the hadronic phase plays a more substantial role in the development of $v_2$ than it does $v_3$.
In all cases the reduction in flow coefficients is substantial, of order 50\% for $v_2$ and even larger for $v_3$ where there is a complete collapse of 
triangular anisotropies without partonic scattering. 

\begin{figure}
\centering
\includegraphics[scale = 0.4]{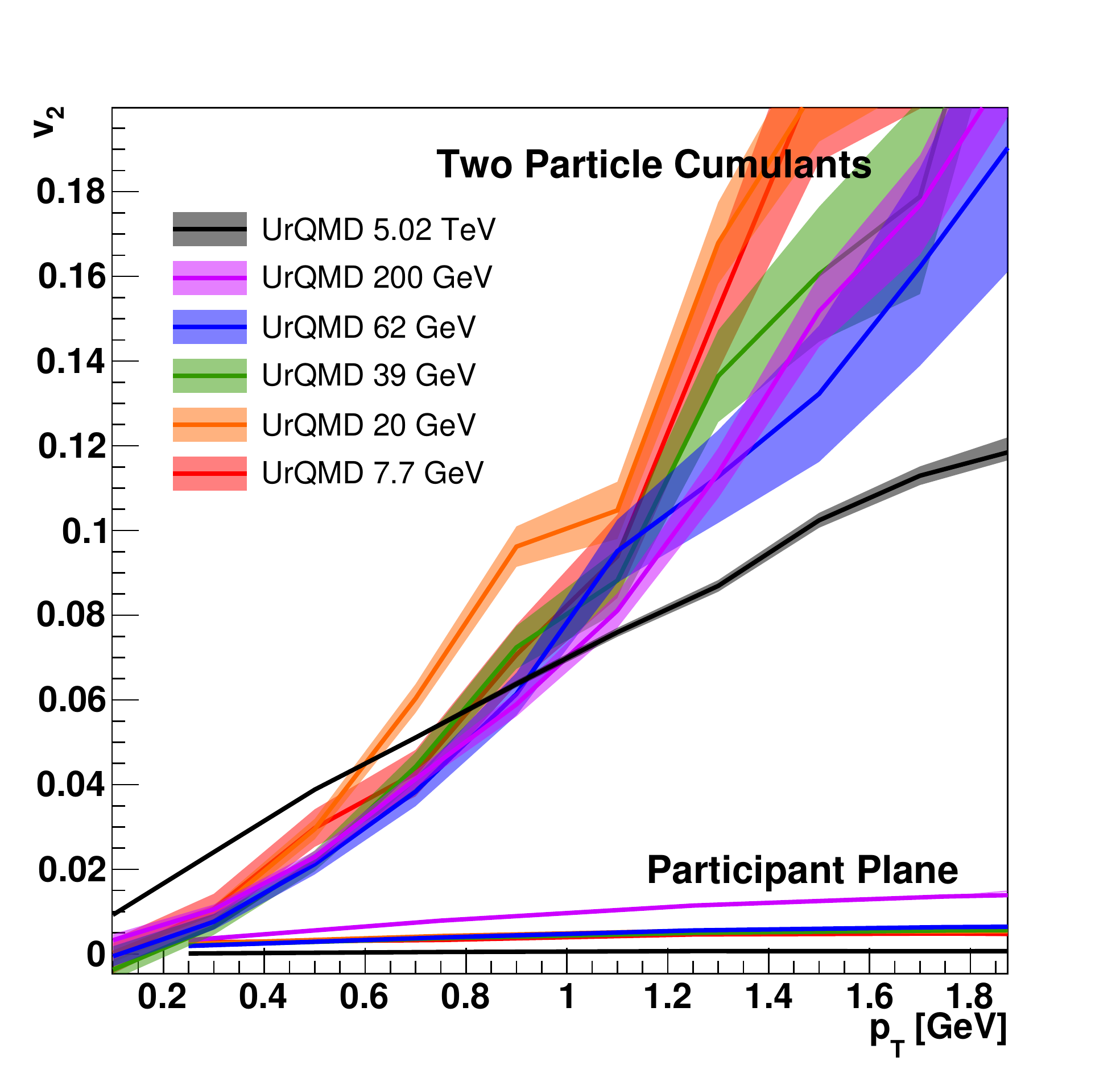}
\caption{Transverse momentum dependence of $v_2$ from the UrQMD model at variety of collision energies, computed with two-particle cumulants and with respect to the participant plane.}
\label{fig_urqmd_comparison}
\end{figure}

Additionally, the purely hadronic \textsc{urqmd} results for $v_2$ are shown in black in Figure ~\ref{fig_v2_all}. The \textsc{urqmd} results exhibit an extremely small $v_2$ that rises modestly with \pt.   Both \textsc{ampt} without partonic scattering and \textsc{urqmd} results fail to reproduce the experimental data in \dau collisions at 200 GeV.   It is notable that these two calculations with hadronic rescattering only give quite different $v_2$ values.
This means that the details of initial hadronic production play a large role in addition to the hadronic rescattering itself.

One hypothesis is that below some size and initial energy deposition, the system can be described
as purely hadronic.    If that regime is reached in lower collision energy \dau reactions, one might expect a significant decrease in the $v_2$ and a 
collapse of the $v_3$ in future lower energy \dau experimental results.    This will be an important test of where partonic effects are no longer important.


Lastly, it is striking that the \textsc{urqmd} $v_2$ results relative to the initial event plane are so small.   In Ref.~\cite{Zhou:2015iba}, they use
two-particle cumulants and find a substantial ``flow'' signal, though they attribute some of the effect to non-flow effects in comparing pseudorapidity
gaps of $|\Delta \eta| > 0.2$ and $|\Delta \eta| > 0.8$.    Shown in Figure~\ref{fig_urqmd_comparison} are our calculation results for the \textsc{urqmd}
participant event plane results and the two-particle cumulant results (in our case with $|\Delta \eta| > 1.6$) in \dau collisions at various energies.   
The two-particle cumulant results
are dramatically larger, have a very steep rise with \pt, and are actually largest for the lowest \dau collision energy.   All of these observations
indicate a complete dominance of non-flow contributions, and thus warrant future caution in the use of two-particle cumulants alone.


\section{\label{sec:summary}Summary}
Recent analyses of RHIC and LHC data have opened the door to the possibility of QGP formation in small systems, a regime where it had been previously given limited consideration. Although the role of initial geometry has been established, a beam energy scan would provide insight into the role of pre-equilibrium dynamics, partonic interactions, and hadronic interactions in the development of flow in these systems.

We find that both hydrodynamics (\textsc{sonic} and \textsc{supersonic}) and \textsc{ampt} predict a substantial $v_2$ and $v_3$ at all collision energies under consideration.   
Within hydrodynamics there is still a significant summed space-time volume of QGP even at the lowest energies, indicating the absence of a sharp turn off in the translation of initial state anisotropies to the final state in small systems.
In \textsc{ampt}, varying \sqsn has only a modest impact on the dynamics of partonic scattering.   Results from \textsc{ampt} without partonic scattering
and the purely hadronic \textsc{urqmd} indicate that if at lower energies the system is purely hadronic, one would expect a significant drop in $v_2$ 
observed via a small system beam energy scan.
New experimental data from the small system beam energy scan at RHIC can shed new light on the required dynamics in these collisions with shorter and shorter lifetimes.

\begin{acknowledgments}
We gratefully acknowledge Paul Romatschke for providing us with the \textsc{sonic} and \textsc{supersonic} curves. We also acknowledge useful discussions on \textsc{urqmd} with Steffen Bass, Hannah Petersen, Huichao Song, and You Zhou; and on \textsc{ampt} with Zi-Wei Lin. We acknowledge funding from the Division of Nuclear
Physics of the U.S. Department of Energy under Grant
No. DE-FG02-00ER41152.
\end{acknowledgments}

\bibliography{references}

\end{document}